\documentclass[final,5p,times,twocolumn]{elsarticle}
\usepackage{amssymb}
\usepackage{amsmath}
\usepackage{booktabs}
\usepackage{tabularx}
\usepackage{placeins}
\usepackage{hyperref}
\usepackage{caption}
\usepackage{bm}
\usepackage[most]{tcolorbox}
\newtcolorbox[auto counter]{forwardmodelbox}[2][]{
  colback=gray!5, colframe=black, boxrule=1pt,
  title=Box~\thetcbcounter: #2,
  #1}

\newtcolorbox[auto counter]{glossarybox}[2][]{
  colback=blue!5, colframe=blue!75!black, boxrule=1pt,
  title=Box~\thetcbcounter: #2,
  #1
}

\journal{Current Opinion in Structural Biology}
\date{\today}

\begin{document}

\begin{frontmatter}

\title{Integrative modelling of biomolecular dynamics}

\author[1]{Daria Gusew\fnref{fn1}}
\author[1]{Carl G. Henning Hansen\fnref{fn1}}
\author[1]{Kresten Lindorff-Larsen\corref{cor1}}
\ead{lindorff@bio.ku.dk}

\cortext[cor1]{Corresponding author}
\fntext[fn1]{These authors contributed equally to this work. The order of the authors were randomly determined with an awk script}

\affiliation[1]{organization={Structural Biology and NMR Laboratory \& the Linderstr{\o}m-Lang Centre for Protein Science, Department of Biology, University of Copenhagen},
addressline={Ole Maal{\o}es Vej 5},
postcode={2200},
postcodesep={},
city={Copenhagen},
country={Denmark}}

\begin{abstract}
Much of our mechanistic understanding of the functions of biological macromolecules is based on static structural experiments, which can be modelled either as single structures or conformational ensembles. While these provide us with invaluable insights, they do not directly reveal that molecules are inherently dynamic. Advances in time-dependent and time-resolved experimental methods have made it possible to capture the dynamics of biomolecules at increasingly higher spatial and temporal resolutions. To complement these, computational models can represent the structural and dynamical behaviour of biomolecules at atomistic resolution and femtosecond timescale, and are therefore useful to interpret these experiments. Here, we review the progress in integrating simulations with dynamical experiments, focusing on the combination of simulations with time-resolved and time-dependent experimental data.
\end{abstract}

\end{frontmatter}

\section{Introduction}
Understanding the functions of biomolecules is important, as these play essential roles across all major life processes, such as enzymatic catalysis \cite{lee_2025}, allosteric regulation \cite{astore_2024}, molecular recognition \cite{holehouse_2024}, protein ligand interactions \cite{boehr_2009}, and other biological phenomena. Many biomolecules are highly dynamic, exchanging between multiple structural states. Often their biological function depends on both their conformational states and the rate of transition between these states. Therefore, fully characterising a biomolecule requires consideration of both structural and kinetic aspects.

Experimental techniques, including X-ray/neutron diffraction and small-angle scattering (SAXS/SANS), nuclear magnetic resonance (NMR) spectroscopy, cryogenic-sample electron microscopy (cryo-EM), Hydrogen-Deuterium Exchange Mass Spectroscopy (HDX-MS), and others, have advanced our understanding of biological processes by providing detailed structural information. These techniques are often interpreted as static, i.e. containing only information on the conformational properties and no direct time-scale information. We here refer to them as  
\emph{static experiments} (Glossary; Table~\ref{tab:glossary}). They are frequently modelled in terms of either a single structure or a conformational ensemble.

While these experiments are the backbone of structural biology, the dynamical details of biomolecules, such as transiently populated states, the timescales of conformational changes, and binding pathways, are difficult to infer from them. 
To address this, the field has developed dynamical experiments such as time-resolved X-ray crystallography and scattering, time-resolved and time-dependent spectroscopies, and various NMR relaxation experiments \cite{van2015integrative}. 
With the term \emph{dynamical experimental data} we here refer to experiments that either explicitly measure changes in an observable over time or where the observable depends on the underlying kinetics. These techniques probe both structural and temporal information, offering more direct insight into the dynamics of biomolecules.

\begin{figure}[htbp]
\centering
    \includegraphics[width=1\textwidth]{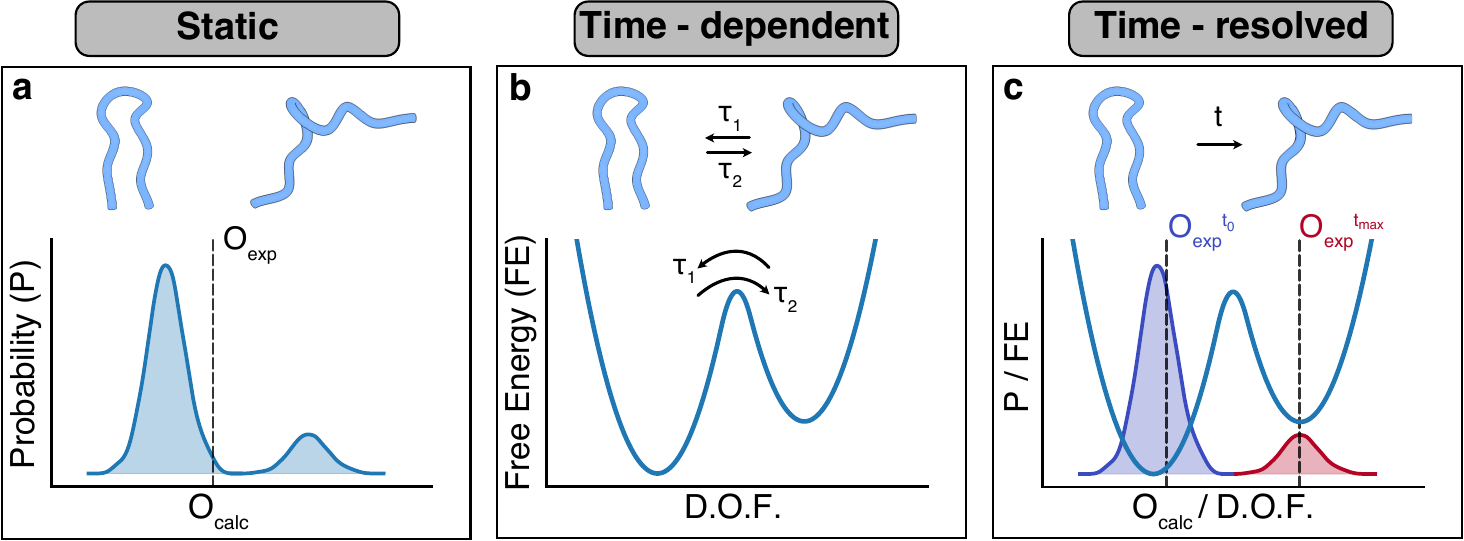}
    \caption{Comparison of a schematic example showing static, time-dependent, and time-resolved experiments illustrated by a protein folding process. (a) A static experiment measuring the observable O$_{\text{exp}}$ is shown, which can be modelled as a distribution of simulated values, O$_{\text{calc}}$, representing a conformational ensemble of folded and unfolded states. (b) Shows a time-dependent experiment, where the equilibrium dynamics of reversible folding gives rise to measured transition times $\tau_1$ and $\tau_2$. These can be modelled as equilibrium dynamics, illustrated by a free energy (FE) surface along a chosen degree of freedom (D.O.F.) (c) A time-resolved experiment probes a non-equilibrium process, where the system begins at $t_{0}$ in the folded state. During the observation time $t$ the protein unfolds until $t_{\text{max}}$. At each time point, a distinct ensemble average, O$_{\text{exp}}$, can be observed, reflecting the proteins changing structure. This evolution can be modelled as distributions of O$_{\text{calc}}$ at each time point. These are shown together with a FE surface.}
    \label{fig:main_fig1}
\end{figure}

For dynamic experiments, we differentiate between \emph{time-dependent} and \emph{time-resolved} observables (Fig.~\ref{fig:main_fig1}). A \emph{time-dependent} observable often depends on a time-correlation function of the system; examples include nanosecond fluorescence correlation spectroscopy, fast-timescale NMR measurements (for example, measurements of backbone motions via R$_{1}$, R$_{2}$ and heteronuclear Nuclear Overhauser Effect (NOE) spin relaxation), or slower motions probed for example via  Carr-Purcell-Meiboom-Gill (CPMG) and R$_{1\rho}^{ex}$ measurements. Time-dependent experiments are most often performed at equilibrium.

In contrast, \emph{time-resolved} experiments typically probe dynamical properties as a series of `static' experiments, where each such experiment reports on the conformational ensemble at a single time point. Time-resolved experiments are generally performed out of equilibrium after a perturbation to the system (e.g. temperature-jump SAXS or time-resolved X-ray crystallography following, for example, triggering via a laser pulse).

We note that some time-dependent observables are often interpreted without explicitly including the effects of the dynamics. For example, while NOE and single-molecule F{\"o}rster resonance energy transfer (smFRET) experiments are sensitive to the timescales of molecular motions, they are often interpreted in a static framework. NOEs are typically treated as $r^{-6}$ or $r^{-3}$ ensemble-averaged interatomic distances, capturing only an approximation of fast-timescale dynamics \cite{vogeli2014nuclear}. Similarly, smFRET experiments are often interpreted in terms of time-averaged distances rather than by explicit treatment of the underlying dynamics \cite{nuesch2025accuracy}. 

Dynamical experiments can probe the motions of molecular processes more directly, but share many of the same limitations as their static counterparts, including limited resolution, random and systematic measurement errors and highly averaged observables. 
Simulations can help deconvolute the measured structural and dynamical behaviour with an atomistic spatial and femtosecond temporal resolution. 
Commonly used computational methods are molecular dynamics (MD) or Monte Carlo simulations, where conformations are sampled based on a force field that describes bonded and non-bonded interactions between atoms or coarse-grained (CG) beads. While these conformations could be generated in multiple ways, e.g via deep learning models \cite{Aranganathan_2025}, we will focus on methods that use MD simulations.

Experiments and simulations can provide complementary information \cite{nasica_labouze_2015, Bernetti_2023, bozovic_real-time_2020}, but notable differences usually still exist between biophysical experimental data and the predictions from molecular simulations. These deviations typically arise from imperfect force fields, insufficient sampling or inaccurate forward models \cite{orioli_2020}. In this context, a forward model describes how a simulation relates to experimental data. It is a model that calculates an experimental observable from either a structure or a trajectory. Most experimental observables are calculated based on atomic coordinates; in the case of conformational ensembles the experimental observable is calculated from each configuration and averaged subsequently. This is inherently only an approximation that neglects dynamic effects unless explicitly included.

There already exist several methods to combine experimental data with simulations \cite{ravera2016critical,cesari_2018,orioli_2020} including for example Metainference \cite{bonomi_2016}, Bayesian Inference of Ensembles (bioEN) \cite{hummer_2015} and the related Bayesian/Maximum Entropy reweighting (BME) \cite{bottaro_2020}. These methods are often called \emph{ensemble refinement} as an umbrella of integrative methods that seek to interpret averaged experimental measurements as an ensemble of
structures. Typically, these methods are applied to static data and result in ensembles that represent the conformational heterogeneity but not explicitly the dynamics. Here, we focus on methods for interpreting time-resolved and time-dependent experiments and suggest possible directions for the future.

\section{Integrating molecular simulations and experimental data}
The process of integrating experimental data with simulations typically requires four components: (1)~A single or multiple experiments that report on static or dynamical observables, (2)~a converged simulation of the system, ideally under the same conditions as the experiment, (3)~a forward model, and (4)~a statistical method for integrating experiment and simulation. Steps (2) and (3) can be carried out separately, so-called integrating \emph{a posteriori}, or together, called integrating \emph{on-the-fly} \cite{orioli_2020}.
We stress that the last three components all present challenges in practice and are the subject of ongoing research. 
Informally, a simulation is said to converge when it samples the relevant states accessible within the time scale of the experiment in a way that enables reasonable error estimates \cite{van_gunsteren_2018}. With the notable exception of some specialised hardware architectures \cite{shaw2021anton} capable of simulating proteins at the millisecond timescale, most all-atom MD simulations of biological molecules cannot currently simulate beyond a continuous runtime of up to tens of microseconds. This is in contrast to the dynamics, which can go all the way from femtoseonds to days. Thus, MD simulations are often constrained by the accessible timescales. This is known as the \emph{time-scale problem}, which refers to the fact that many biologically important phenomena (e.g. protein folding, conformational changes or ligand binding) occur at timescales inaccessible directly by atomistic MD simulations. In relation to this is the \emph{sampling problem}, which is the inability of atomistic MD simulations to explore the relevant configurational space and sample important slow motions of many biomolecular systems \cite{van_gunsteren_2018}. In response to this, the field has proposed many solutions, two of which we shall cover briefly here: Enhanced sampling and coarse graining.

CG models mitigate the time-scale problem by eliminating less important degrees of freedom, while preserving the essential features of the system, making it possible to simulate larger systems for longer times by reducing the computational cost.
Essentially, a cluster of atoms is represented by a single pseudoatom (referred to as a CG bead), with the number of atoms per bead defining the resolution. The main challenge is to construct a force field that accurately captures physical interactions between beads. CG models have been used to investigate, for example, protein folding dynamics and conformational dynamics of intrinsically disordered proteins (IDPs) \cite{ingolfsson_2014}. While some CG models have shown high accuracy in reproducing experimental measurements, their success depends heavily on the quality of parametrisation, including both CG mapping and the functional form and parameters in the force-field.

The idea behind enhanced sampling is to change the dynamics of an MD simulation in such a way that sampling is sped up. Often, this change in the dynamics is achieved by applying a potential (often called a bias) that enables the dynamics to cross potentially very high (free) energy barriers. The physical property that is recovered afterwards is most often a free energy surface, along a set of specifically defined degrees of freedom representing the slow motions of the system (referred to as a collective variable). The tradeoff is that some amount of prior knowledge of the physical process under investigation is generally required. There are many approaches to do this including metadynamics, adaptive biasing force, umbrella sampling, replica exchange, and more \cite{Henin_Lelievre_Shirts_Valsson_Delemotte_2022}. 

Both coarse-graining and enhanced sampling have found wide application, but come with drawbacks when modelling dynamical experiments \cite{ray_2023, wolf_2023}. 
Since both approaches often change the force field to enable faster sampling, they do not directly model realistic physical timescales. This makes direct comparison to dynamical experiments difficult \cite{jin_2023}; we call this the \emph{timescale reconstruction problem}. Other approaches exist that do not directly change the force field, but accelerate sampling in other ways, are milestoning, transition path sampling, and weighted ensemble \cite{Henin_Lelievre_Shirts_Valsson_Delemotte_2022}. 

\subsection*{Bayesian/maximum Entropy Reweighting}
As discussed above, a number of different integrative approaches exist to combine static experiments with simulations. In the simplest case, one could refine a molecular ensemble model by selecting those structures ($x_i$) from a conformational ensemble ($\bm{x}$) whose observable $O_{\text{calc}}(x_i)$, calculated through the forward model, agrees well with the experimental measurement $O_{\text{exp}}$, while discarding the rest. This approach would, however, ignore the underlying dynamics and can lead to conclusions that are unrepresentative of the distribution of configurations behind $O_{\text{exp}}$. One way to solve this problem is to provide prior information that constrains the model to behave in accordance with some expected behaviour. 

One popular approach to interpret static experiments in terms of conformational ensembles is Bayesian/maximum entropy reweighting, where the goal is to change the relative populations of each configuration in a way that improves agreement with experimental data \cite{hummer_2015,orioli_2020}. When an ensemble $\bm{x}$ is generated, a prior relative weight $w^0_i \in \bm{w}^0$ is assigned to each $x_i$ and the calculated observable becomes $O_{\text{calc}}=\sum_{i}w_iO_{\text{calc}}(x_i)$. In the case of unbiased MD, all weights are identical. Agreement with the experiment can be expressed as the likelihood 
\begin{align}
\mathcal{L}(\bm{w}) = p(\bm{O}_{\text{exp}}|\bm{O}_{\text{calc}}(\bm{x},\bm{w})).
\end{align}
As regularisation, the relative entropy (or negative Kullback-Leibler divergence) $S_{\text{rel}}(\bm{w}| \bm{w}^0)$ is a measure of the difference between a given set of weights $\bm{w}$ and the prior weights $\bm{w}^0$. A Bayesian or maximum entropy (MaxEnt) solution gives the least biased distribution given the data. By minimising a functional such as 
\begin{align}
    \Gamma(\bm{w}|\bm{w}^0) = -\log \mathcal{L}(\bm{w}) - \theta S(\bm{w}|\bm{w}^0),
\end{align}
one can optimise agreement with the experiment, while penalising straying too far from the original ensemble. Here, $\theta$ is a hyperparameter that quantifies the trust in the prior and is required when the experimental errors are not well estimated, which is often the case. Modelling dynamical experiments using this type of approach can be challenging, as the form of this functional is not as straightforward. 

In the following, different methods for the integration between dynamical experiments and simulations for unbiased MD simulations and enhanced sampling will be explained. A short list of important technical terms related to integrative methods can be found in the glossary (Table~\ref{tab:glossary}). For a more technical review of integrative methods, the reader is referred to previous literature \cite{hummer_2015,ravera2016critical,cesari_2018,orioli_2020, bottaro_2020}. 

\section{Integrative Methods using Unbiased Simulations}
\subsection*{Methods for Conventional Molecular Dynamics}
Some experiments report on timescales fast enough that it is possible to reversibly sample transitions via conventional MD. An example of such an experiment is backbone or side-chain NMR spin relaxation, which probes motions on ps-to-ns timescales \cite{palmer_nmr_2004}. A transition event generated through MD is called a transition path and is an ordered sequence of frames, $\gamma = (x_t, x_{t+1}...x_T)$. From an ensemble of such paths, any dynamical observable can be generated. This is very powerful when combined with path sampling algorithms, where transition paths can be sampled efficiently, even for rare events \cite{dellago_2006}.

A method called continuum path ensemble maximum caliber (CoPE-MaxCal) \cite{brotzakis_2021} was developed as an integrative method to model experimentally determined rate constants using both transition paths and non-reactive paths. The method works by reweighting the path ensemble to match an experimental rate constant, while minimally perturbing the path distribution. This is done using a combination of MaxEnt and by maximising the path entropy (referred to as the caliber).
The caliber is the dynamical equivalent to entropy and quantifies the similarity between two path ensembles \cite{Jaynes_1980}, and such methods are often referred to as Maximum Caliber (MaxCal) methods. 
Maximising the caliber between the sampled path ensemble and an optimised one ensures a minimally perturbed ensemble. 
CoPE-MaxCal was illustrated on multiple test systems, a 2D free energy landscape toy model and all-atom MD simulations of the folding dynamics of the designed 10-residue peptide chignolin. For this, a more accurate melting temperature after path reweighting as well as a more native-like transition-state ensemble was predicted.

To model NMR relaxation rates for intrinsically disordered proteins, a method called ABSURD (Average Block Selection Using Relaxation Data) was developed \cite{salvi_2016}. This method uses an ensemble of path segments (MD simulation blocks) to model a measured relaxation rate. A conventional MD simulation is run and split into a number of blocks of fixed size. These blocks are used to calculate relaxation observables via correlation functions that capture the dynamics relevant to the NMR spin-relaxation experiments. The relative weight of each block is changed to minimise the difference between the experimental and calculated relaxation rates across different magnetic fields and relaxation rate types. An extension of this framework, called ABSURDer \cite{kummerer_2021}, incorporates a caliber-like restraint when optimising the agreement with experiment. The method was demonstrated on deuterium relaxation measurements probing side-chain motions in T4 lysozyme \cite{kummerer_2021}.

For time-resolved experimental data, a MaxCal method was introduced in which several replica simulations are run simultaneously and combined on the fly with time-resolved ensemble-averaged experimental data \cite{capelli_2018}. As the time-resolved data represents relaxation towards an equilibrium, the simulations are started from an initial state and are then allowed to relax. It was shown that adding a time-dependent harmonic potential to these simulations is equivalent to reweighting the path ensemble to maximise a likelihood conditioned on maximising the caliber. The method is demonstrated on synthetic experimental time-resolved data of various types including SAXS for protein test systems.
Other notable applications of MaxCal are in force field parametrisation from dynamical data \cite{bolhuis_2023} 
and to refine simulations based on the milestoning framework \cite{ji_2025}.

\subsection*{Methods for Markov State Models}
For experiments beyond the timescales accessible to conventional MD, many shorter parallel unbiased simulations can be combined to simulate transition paths and analysed using Markov State Models (MSMs) \cite{husic_2018}. 
An MSM is used to model the dynamical properties of a system, usually in thermodynamic equilibrium, where the entire configurational space is divided into $n$ discrete conformational states $S_{i},...,S_{n}$ and their dynamics is described by an $n\times n$ matrix, called the transition matrix $T^{n\times n}$ estimated from the simulation trajectories $x$. 
Each element represents a probability $p_{ij}$ of the system transitioning from state $i$ to $j$ after a lag time $\tau$, and is defined as 
\begin{align}
    T^{n\times n}\equiv p_{ij}(\tau) \text{ with  } p_{ij} = \text{Prob}(x_{t + \tau} \in S_{j} | x_{t} \in S_{i}).
\end{align}
MSMs are useful in mitigating the time-scale and sampling problems, as the slow dynamics of a system can be reconstructed from many shorter and parallel MD simulations. This reconstruction estimates the time scales of the simulations and thus avoids the time-scale reconstruction problem. This enables extrapolation to effective timescales longer than the length of an individual MD simulation.   
MSMs exploit the fact that many biomolecular systems exhibit Markovian behaviour (i.e. memoryless and stochastic) dynamics above a certain lag time $\tau$.  
In that case, the eigendecomposition of the transition matrix yields a set of eigenvectors and eigenvalues that describe the dynamics of the system, where the eigenvector belonging to the largest eigenvalue represents the stationary distribution. The eigenvalues give information on molecular timescales, while the eigenfunctions describe structural changes of the system. 
In practice, the transition matrix is generated by assigning a set of discrete states to the MD simulations via a chosen clustering algorithm and careful featurization of the input coordinates or collective variables from the MD trajectories. $\tau$ is then chosen as the smallest lag time for which the transitions appear to be Markovian \cite{husic_2018}. 

Several methods exist to integrate experimental measurements into the estimation of an MSM, ensuring that the resulting model reflects experimental observables accurately. Augmented Markov State Models (AMMs) combine static experimental observables and MD simulations with MaxEnt during the estimation of an MSM model \cite{olsson_2017}. Using Ubiquitin as an example, it was shown that differences between equilibrium distributions from different force fields can be decreased via AMMs that were augmented with NMR observables such as NMR scalar and residual dipolar couplings. Integrating only static experimental data with simulations in the AMM framework also improved agreement with dynamic observables, suggesting that improved kinetics came as a byproduct of improved equilibrium behaviour. 
Building on this approach, dynamic Augmented Markov State Models (dynAMMo) \cite{kolloff_2023} include not only static but also dynamical data by combining  AMMs with dynamical experimental constraints. This approach was shown to work even when rare events are not sampled in MD simulations, since experimental data is used to account for these processes. The method is demonstrated on two benchmark systems, as well as on the conformational dynamics in BPTI by integration of CPMG relaxation data with atomistic MD simulations.

In another integrative method, the MSM is biased by enforcing kinetic constraints using a tunable bias parameter that does not prescribe the type of experimental data used \cite{rudzinski_2016}. Caliber Corrected Markov Modeling (C$_{2}$M$_{2}$) augments an MSM based on MaxCal with position-dependent diffusion coefficients \cite{dixit_2018}. 

\section{Integrative Methods using Enhanced Sampling}
\subsection*{Reconstructing Dynamics from Biased Simulations}
Since the MSM framework builds on unbiased sampling of inter-state transitions, it becomes difficult to use when the energy barriers between metastable states are high or when the transitions are very slow. Many enhanced sampling methods can overcome high free energy barriers and obtain a correct free energy surface, but come with the cost of altering the physical timescales of the simulation.

To perform integrative modelling on systems where the use of MSMs is not practical, we developed a method called time-resolved Bayesian/Maximum Entropy reweighting (trBME) for using free energy surfaces from enhanced sampling to model time-resolved experiments \cite{hansen_2025}. We use the popular assumption that the dynamics along a good collective variable can be described by an overdamped Langevin equation \cite{shaw_2010, wolf_2018}, even though it is only strictly true for a perfect reaction coordinate \cite{peters_2017}. Even when this assumption does not strictly hold, the inclusion of experimental data is expected to compensate for this. In trBME, the conformational dynamics is modelled with Langevin dynamics on a low-dimensional free energy surface estimated from enhanced sampling MD simulations. 
By scaling the Langevin time scale to match the experimental time, we can simulate dynamics on the free energy surface. To do this, a set of time-dependent probability densities are generated  that serve as a  \emph{dynamical prior}, which are then reweighted at each timepoint using MaxEnt. The method is illustrated on synthetic temperature-jump time-resolved SAXS data of unfolding of bovine serum albumin \cite{hansen_2025}.

The idea of using low-dimensional models to interpret experimental data can also be applied to time-dependent experiments, such as NMR relaxation dispersion experiments. 
In \cite{daffern_2022}, chemical exchange is modelled via transitions between conformational states, assuming Langevin dynamics on a lower-dimensional energy surface. Subsequently, transitions between conformational states were used to evaluate chemical shift autocorrelation functions to analyse NMR CPMG relaxation dispersion profiles. Building on this method and by using the same assumptions as in trBME, chemical exchange dynamics of a protein could possibly be modelled on a one-dimensional free energy profile extracted from enhanced-sampling simulations and using experimental data to reconstruct the timescales by matching experiments. 

\section{Discussion and Future Perspectives}
\begin{figure}[htbp]
    \centering
    \includegraphics[width=1\textwidth]{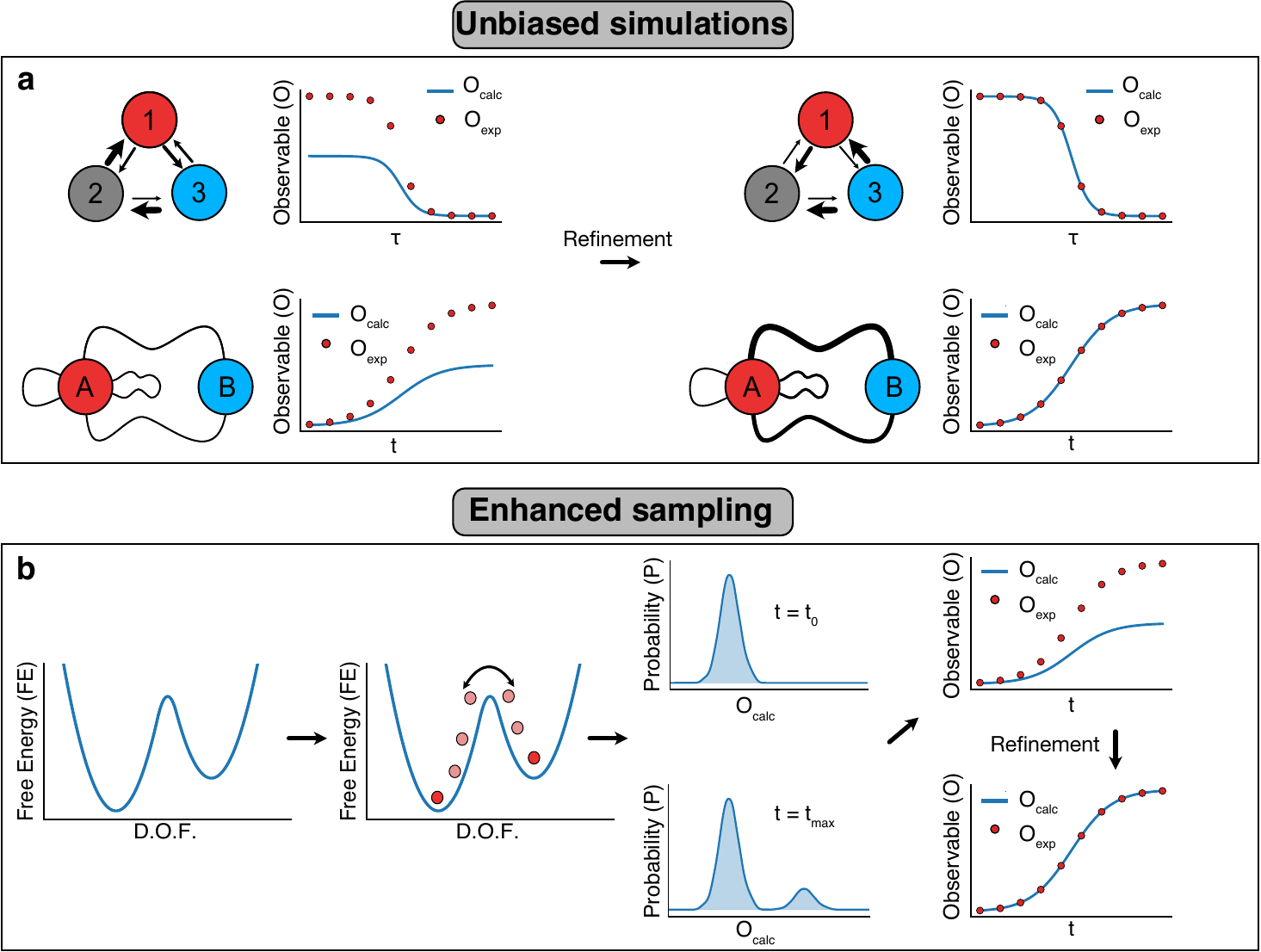}
    \caption{Integration of dynamical experimental data using different types of molecular dynamics simulations. (a) For unbiased simulations, there are two main approaches integration via path reweighting (bottom) and via Markov State Models (top). Dynamic Augmented Markov Model (dynAMMo) use dynamical experimental data to refine a Markov State Model and subsequently improve the estimation of dynamic observables. Maximum Caliber (MaxCal) methods reweight a path ensemble to better match dynamic experimental data. (b) Illustration of using enhanced sampling simulations via the time-resolved Bayesian/Maximum Entropy (trBME) approach: A free energy (FE) surface is extracted from an enhanced-sampling molecular dynamics simulation. The non-equilibrium experimental process is modelled using Langevin dynamics on this FE surface and then refined to allow the reconstruction of dynamics from biased simulations that are consistent with experimental data.}
    \label{fig:main_fig2}
\end{figure}

We have here reviewed methods to integrate experimental data with simulations to study the dynamics of biomolecules, focusing on integrating dynamical experimental data with MD simulations. With dynamical data we refer to either time-dependent or time-resolved observations. By time-dependent we mean that the individual measurements depend both on the conformational ensemble and the associated timescales, whereas time-resolved experiments typically are performed as a series of `static snapshots' of the system. We have discussed two main classes of methods, those that are compatible with unbiased simulations (Fig.~\ref{fig:main_fig2}a) and those that are compatible with enhanced sampling (Fig.~\ref{fig:main_fig2}b). The methods that utilise unbiased simulations are further distinguished by whether they utilise an MSM framework or a path space framework.

While new methods are being developed to integrate diverse experimental data and to improve existing approaches, the number of computational methods to interpret dynamical experiments is still limited.
Beyond the technical details of how to integrate experiments and simulations, one major challenge is determining which representation of the experimental data to use in the integration. While it can be argued that working with the least processed form of the experiments introduces less bias, this may not always be feasible or even optimal. As one example, NMR spin relaxation measurements could be represented either as `raw' free induction decay NMR data, as a series of peak intensities, as fitted relaxation times or even as generalized order parameters in a fitted model. Depending on the choice, different amounts of dynamical information may be extracted from the experiment.

Since the application of MD is in many cases hindered by the sampling problem, new integrative approaches should focus on this challenge. This could be methods that support the use of CG force fields or enhanced sampling. For the use of enhanced sampling, a field of promising ongoing research is dynamical reweighting. This is a series of methods that seek to recover dynamics from biased simulations based on enhanced sampling \cite{keller_2024}, and may enable extraction of long-time scale dynamics from simulations. 

Another future direction that could improve MD for integrative modelling is the use of deep learning methods to efficiently and quickly generate trajectories and structural ensembles. Much work is currently being directed towards these methods and they will likely play a role in future modelling of dynamical experiments. Some machine learning approaches aim to enhance the sampling of the actual molecular dynamics \cite{schreiner2023implicit,klein2023timewarp}; other methods could be used to generate starting points for more conventional MD simulations if the two are thermodynamically consistent \cite{chennakesavalu2023ensuring,lewis2025scalable}.

Finally, an important area that we did not cover in this review is force-field improvement. Dynamical experiments could improve the predictive accuracy of molecular force fields, and progress has already been made in incorporating such data into their optimisation \cite{aliev2014motional,bolhuis_2023, kummerer_fitting_2023}. It is a natural next step to go from integrative methods to systematically improving our physical models.

In summary, dynamical experiments provide a unique opportunity to expand integrative structural biology further into studies of biomolecular dynamics. Much progress has already been made on the experimental side, and computational models are increasingly focused not only on comparing simulations and dynamical experiments, but also on directly integrating the two. We envisage that a tight integration and joint development of experiments, theory, machine learning and simulations will continue to push the field forward.

\section{Acknowledgements}
We acknowledge support by the European Union (ERC, DynaPLIX, SyG-2022 101071843). Views and opinions expressed are however those of the authors only and do not necessarily reflect those of the European Union or the European Research Council. Neither the European Union nor the granting authority can be held responsible for them.

\section{Declaration of competing interest}
K.L.-L. holds stock options in, receives sponsored research from, and is a consultant for Peptone Ltd. The remaining authors declare no competing interests.

\bibliographystyle{elsarticle-num} 
\bibliography{references}

\clearpage
\begin{table}[htbp]
\centering
\caption{Glossary of technical terms.}
\begin{tabularx}{\textwidth}{XX}  
    \hline
    \textbf{Term} & \textbf{Definition} \\ \hline
    Conformational Ensembles & A collection of three-dimensional structures and their weights sampled from a distribution of conformations a molecule can adopt. \\
    Dynamical Experiment & An umbrella term for both time-dependent and time-resolved experimental data.\\
    Ensemble Refinement  & Methods that seek to model an, often static, experiment as a conformational ensemble. This involves changing the  ensemble such that it is in agreement with some experimental data.\\
    Force Field & A set of functions and parameters describing the particles and their interactions in a simulation. \\ 
    Forward Model & A method that calculates a proposed experimental signal from molecular coordinates.  \\ 
    Integrative method & A method that uses one or more pieces of experimental data and a computational method to construct a combined model of a physical system.  \\
    Sampling problem & It is too computationally expensive to sample all relevant conformational states of a biomolecular system with current simulation methods.\\
    Static Experiment & A static experiment reports on an observable that contains structural information, but is not, to some level of approximation, directly dependent on the time scales of the system. Such experiments can be modelled as single structures or ensembles of structures.\\
    Time-Dependent Experiment & An experiment that reports on a function of multiple points in time, often measured in equilibrium, e.g. NMR relaxation experiments.\\
    Time-Resolved Experiment & An experiment performed out of equilibrium such as T-jump time-resolved SAXS or light-activated time-resolved X-ray crystallography, where the measurement is effectively instantaneous, such that each measurement is a single point in time.\\
    Time-scale problem & The problem that many biomolecular processes are on time-scales too slow to reliably estimate from MD simulations.\\
    Time-scale reconstruction problem & Some MD methods can speed up sampling of biomolecular systems, but come at the cost of losing information on the physical time scales of the system, which means that reconstructing information on the physical times scales is difficult.\\
    \hline
    \label{tab:glossary}
\end{tabularx}
\end{table}

\clearpage

\clearpage
\clearpage

\clearpage

\end{document}